\documentclass[Journal]{IEEEtran}
\usepackage{amsmath,amsthm}
\usepackage{cite}
\usepackage{graphicx}
\usepackage{epstopdf}
\usepackage{color}
\usepackage{amsfonts,amsmath,amssymb}
\usepackage{cite}
\usepackage{graphicx}
\usepackage{url}
\usepackage{bm}
\usepackage{bbm}
\usepackage{amssymb}
\usepackage{fancyhdr}
\begin{document}

\title{\huge{Quantum Communications via Satellite with Photon Subtraction}}

\author{Mingjian He$^1$, Robert Malaney$^1$ and Jonathan Green$^2$
\thanks{Mingjian He and Robert Malaney  (email: r.malaney@unsw.edu.au) are with the School of Electrical Engineering and Telecommunications, the University of New South Wales, Sydney, NSW, Australia. Jonathan Green is with Northrop Grumman Mission Systems, San Diego, California, USA. Approved For Public Release $\#$18-1963; Unlimited Distribution. Dated 9/13/18.}}

\maketitle
\vspace{-15mm}

\begin{abstract}
Non-Gaussian continuous-variable quantum states represent a pivotal resource in many quantum information  protocols. Production of such states can occur through photonic subtraction processes either at the transmitter side prior to sending a state through the channel, or at the receiver side on receipt of a state that has traversed the channel. In the context of quantum protocols implemented over communication channels to and from Low-Earth-Orbit (LEO) satellites it is unclear what photonic subtraction set-up will provide for the best performance. In this work we show that for  a popular version of continuous-variable Quantum Key Distribution (QKD) between terrestrial stations and  LEO  satellites, photon subtraction at the transmitter side is the preferred set-up. Such a result is opposite to that found for fiber-based implementations. Our results have implications for all future space-based missions that seek to take advantage of the opportunities offered by non-Gaussian quantum states.

\end{abstract}

\section{Introduction}

Quantum Communications via satellite offers a paradigm shift in our ability to deploy quantum information protocols over very large scales, e.g. \cite{bedin,Neda1,Neda2, neda_rev}. Propagation through the atmosphere to and from LEO satellites can overcome the scourge of the roughly $100$km limited distance  that plagues point-to-point optical-fiber optical links and free-space-optical links. Indeed, in the past few years great strides have been made  in regard to actual
 deployments of quantum communications via satellites\cite{china1,china2,china3,gers,japan}.
These latter works on satellite-based quantum communications are largely based on the deployment of discrete-variable (DV) quantum information protocols, a technology that is dependent on the production of single-photon states.

Continuous-variable (CV) technology offers a different pathway to the implementation of quantum  information protocols. The main advantage of CV technology over DV technology is that detection can be realized by more reliable, and more efficient  homodyne (or heterodyne) detectors e.g., \cite{CV1, CV2, thesis, Weedbrook2012}. Indeed, it is argued by many that relative to DV detectors,
 CV based-detectors offer the promise of a more pragmatic route to higher  secret key rates for certain QKD protocols, e.g. \cite{pir}.

Currently, no experimental deployment of space-based CV quantum technology has been carried out, but this is expected to change soon (see \cite{neda_rev} for review). CV technologies are largely based around so-called Gaussian states, e.g. \cite{thesis, Weedbrook2012}  - quantum states in which the quasi-probability distribution (the Wigner function) of the electromagnetic-field quadratures follow a Gaussian distribution.
However,
the use of non-Gaussian states in the implementation of CV quantum information protocols has also garnered interest, e.g. \cite{nG-modulation, nG1, nG2, nG-coherent, Neda3}.
Non-Gaussian operations such as photon subtraction (PS) \cite{1st_PSS, 2, telep, 3, 9, Oxford, beijing}  on a mode of an incoming two mode squeezed vacuum (TMSV) state can lead to higher levels of entanglement, potentially higher secret (QKD) key rates, as well as forming a pivotal resource for quantum error correction.

In this work we will focus on single PS as a means to produce non-Gaussian states. We will be specifically focussed on the question as to whether PS at the transmitter offers a better pathway to improved QKD (higher secret key rates) when propagation  between ground stations and LEO satellites is considered. The answer to this question has important implications not only for future space-based implementations of CV-QKD protocols, but also potentially for other space-based quantum information protocols that utilize non-Gaussian states.

The structure of the remainder of this paper is as follows. In Section~II, the nature of the quantum channel between terrestrial stations  and  LEO satellites is described. In Section~III, a model for CV-QKD with PS at the transmitter is described, whilst in Section~IV a system for PS at the receiver is described. In Section~V our performance analysis is described, and in Section VI our simulation results are presented, comparing key rates produced from both systems.

\section{Earth-Satellite Channels}

 We consider the model of  single uplink and single downlink satellite channels in an entanglement-based version of a CV-QKD  protocol.\footnote{Each entanglement-based QKD protocol has an equivalent prepare and measure scheme that will give, in theory, exactly the same results.} Our quantum information carrier will be a pulsed optical beam.
For the uplink, we assume that Alice first prepares a TMSV state ($A_0-B_0$) at a ground station, subsequently sending  one of her modes ($B_0$) to the satellite. For the downlink, the TMSV is prepared on the satellite with $B_0$ being sent to the ground station.

 For optical signals in the uplink channel, the dominant loss mechanism will be beam-wander caused by turbulence in the Earth's atmosphere  \cite{fso}.  Assuming the beam spatially fluctuates around the receiver's  center point, the fading of the signal as a consequence of the beam-wander can be described by a distribution of transmission coefficients (amplitude attenuation) $\eta$. The probability density distribution of these coefficients, $p(\eta)$, can be approximated   by the log-negative Weibull distribution, given by \cite{20} \cite{21}

\begin{equation}\
p\left( \eta  \right) = \frac{{2{L^2}}}{{\sigma _b^2\lambda \eta }}{\left( {2\ln \frac{{{\eta _0}}}{\eta }} \right)^{\left( {\frac{2}{\lambda }} \right) - 1}}\exp \left( { - \frac{{{L^2}}}{{2\sigma _b^2}}{{\left( {2\ln \frac{{{\eta _0}}}{\eta }} \right)}^{\left( {\frac{2}{\lambda }} \right)}}} \right)
\label{f1}
\end{equation}
for $\eta  \in \left[ {0,\,{\eta _0}} \right]$, with $p\left( \eta  \right) = 0$ otherwise.
Here, ${\sigma _b}^2$ is the beam wander variance,
 $\lambda$ is the shape parameter,  $L$ is the scale parameter, and ${\eta _0}$ is the  maximum transmission value. The latter three parameters are given by
 \begin{equation}
\begin{array}{*{20}{l}}
&\lambda  = 8h\frac{{\exp \left( { - 4h} \right){I_1}\left[ {4h} \right]}}{{1 - \exp \left( { - 4h} \right){I_0}\left[ {4h} \right]}}{\left[ {\ln \left( {\frac{{2\eta _0^2}}{{1 - \exp \left( { - 4h} \right){I_0}\left[ {4h} \right]}}} \right)} \right]^{ - 1}},\\
\\
&L = \beta_r{\left[ {\ln \left( {\frac{{2\eta _0^2}}{{1 - \exp \left( { - 4h} \right){I_0}\left[ {4h} \right]}}} \right)} \right]^{ - \left( {{1 \mathord{\left/
 {\vphantom {1 \lambda }} \right.
 \kern-\nulldelimiterspace} \lambda }} \right)}},\\
\\
&{\eta _0}^2 = 1 - \exp \left( { - 2h} \right) ,
\end{array}
\label{f2}
 \end{equation}
where ${I_0}\left[ . \right]$ and ${I_1}\left[ . \right]$ are the modified Bessel functions, and where $h = {\left( {{\beta_r \mathord{\left/
 {\vphantom {a W}} \right.
 \kern-\nulldelimiterspace} W}} \right)^2}$, with $\beta_r$ being the aperture radius and  $W$  the beam-spot radius. Here we set $\beta_r=W=1$ unit length (which for typical configurations is 1 meter).

In the downlink satellite channel diffraction effects are anticipated to dominate. This is largely because  beam-wander in the downlink is relatively suppressed since the beam-width, on entry into the atmosphere from space, is generally broader than the scale of the turbulent eddies \cite{fso}. As such, with well-engineered designs\footnote{This involves properly-dimensioned lenses, use of state-of-the-art adaptive optics, and use of feedback from concurrent classical channel measurements. On the latter measurements we note fluctuations caused by turbulence are in the kHz  range (compared to the Mhz rate of the laser pulses), thus allowing for channel-coefficient measurements to be made dynamically (within the coherence time of the channel) by a ground receiver.} losses in the downlink can be as small as 5-10 dB, compared to the 20-30 dB losses that can be anticipated for well-engineered uplink channels. For simplicity, we model all losses by varying $\sigma_b$.

 To investigate the effect of the PS we mainly consider three schemes. The first scheme is where there is no PS (No-PS).  The second scheme is  PS at the transmitter side (T-PS), where the PS is performed immediately after Alice prepares her TMSV state. The last scheme is  PS at the receiver side (R-PS), where Bob performs the PS after he receives the mode from Alice, but before his homodyne measurement.  We adopt the QKD protocol of \cite{5n}, modified as required for our additional T-PS scheme. Reverse reconciliation at Alice, in which both Alice and Bob undertake  homodyne measurements is always used. We will assume the asymptotic limit in the number of measurements taken.

\section {Photon subtraction at transmitter side }

 \begin{figure}[h]
	\includegraphics[width=0.5\textwidth]{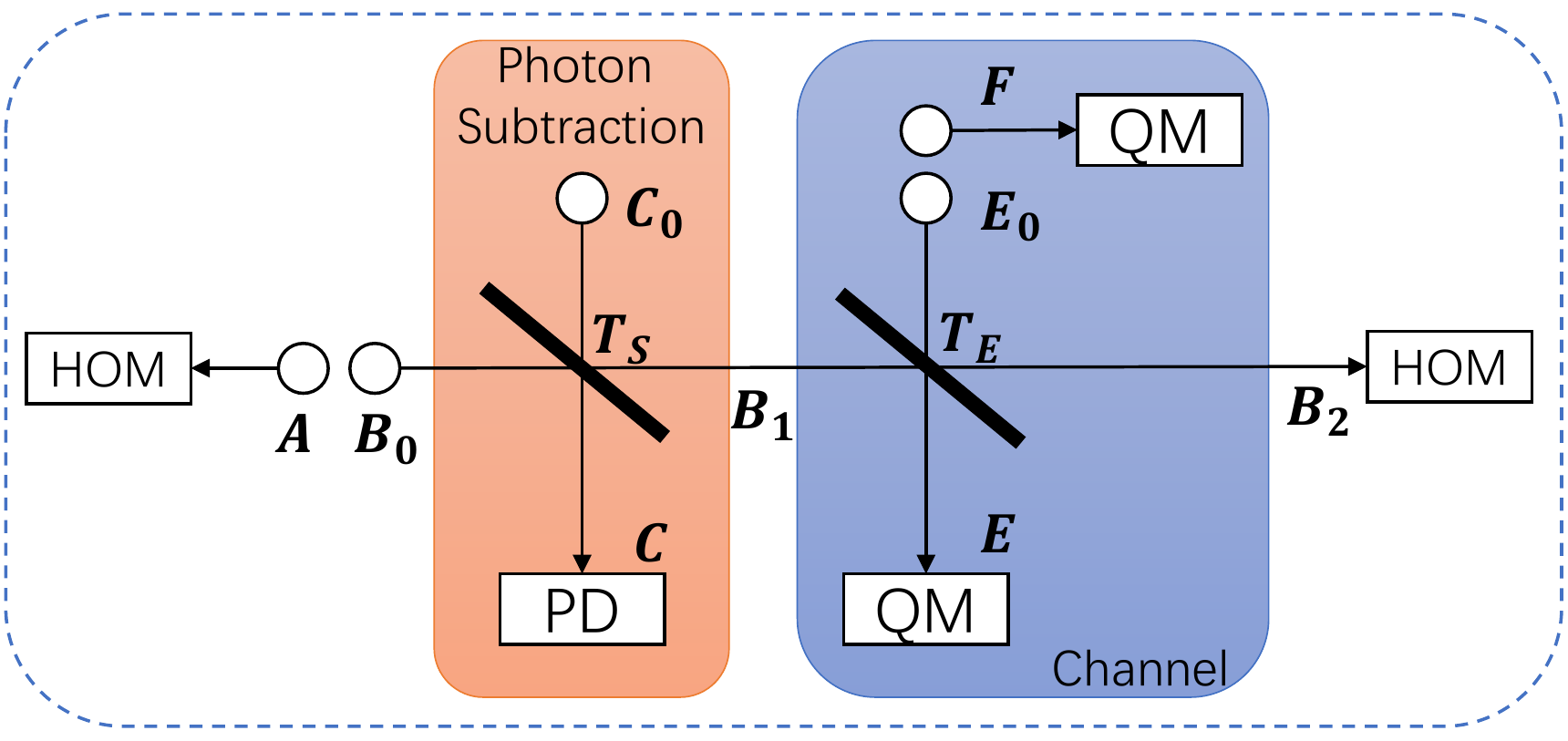}
	\caption{Photon subtraction at transmitter side (T-PS).
 Here Alice (ground station) prepares a TMSV ($A_0-B_0$), sending $B_0$ through a PS process using a beam-splitter with transmissivity $T_S$. The exiting mode $C$ is sent to a photodetector, whilst the exiting $B_1$ is sent to Bob (the satellite). The channel is controlled by Eve using a second beam-splitter with transmissivity $T_E$.
\label{TPS}}
\end{figure}

The system model for the CV-QKD protocol with photon subtraction is illustrated in Fig.~\ref{TPS}.  We assume that Alice first prepares a TMSV $A_0-B_0$ at her ground station (for briefness we just describe the uplink). She then sends one of her modes ($B_0$) through a PS process in which $B_0$ interacts with a mode $C_0$ at a beam-splitter with transmissivity (intensity attenuation) $T_S$. One of the exiting modes ($C$) is sent to a photodetector (PD), whilst the other ($B_1$) is sent to Bob (the satellite). In the following we take mode $C_0$ to be a vacuum state.\footnote{We note that a PS at the transmitter in the context of a somewhat different QKD protocol from that studied here has been investigated for the Earth-satellite channel \cite{Neda5qkd}.}

In this work we assume that Eve performs a collective attack.\footnote{A collective attack is where Eve creates  a series of ancillary  modes with a member from this series independently entangling with each incoming mode sent by Alice. Following Bob's measurements Eve then takes an optimal collective measurement on her series of ancillary modes. In the asymptotic limit, security under collective attacks can be shown to be equivalent to security under coherent attacks (for many protocols) in which Eve's ancillary modes are no longer constrained to interact independently with  Alice's modes.} The channel can then be modeled by Eve feeding one mode,  $E_0$, of a TMSV state ($E_0-F$) prepared by her  into a beam-splitter with transmissivity $T_E$,   with $B_1$ being fed into the other input mode of the beam-splitter. After passing through Eve's beam-splitter, Eve retains the quantum state $F$-$E$, $E$ being one of the output modes of her beam-splitter. The other output mode of the beam-splitter is forwarded to Bob. Setting $T_E={\eta}^2$, we assume that Eve sets $T_E$ so as to follow a probability density function given by equations~(\ref{f1})-(\ref{f2}). Following its traversal through the channel Bob then receives an ``attenuated" version of $B_1$, namely $B_2$.

 Note that PS is not a Gaussian operation, but rather an operation that transforms a Gaussian state into a non-Gaussian state.
Because of this, the state following the PS cannot be fully described by the first and second moment of the quadrature operators $\hat x$
 and $\hat q$ of the electromagnetic field.  As such, a somewhat more complex state description is required relative to that used for quantum protocols based on Gaussian-states. We now describe this more complex quantum state.

 Using the Fock basis, Alice's initial TMSV state ${\left| \psi  \right\rangle _{A{B_0}}}$
 has the form
 \[{\left| \psi  \right\rangle _{A{B_0}}} = \sum\limits_{n = 0}^\infty  {{\alpha _n}} {\left| {n,n} \right\rangle _{A{B_0}}} \ ,\]
 with
$${\alpha _n} = \sqrt {\frac{{{\alpha ^{2n}}}}{{{{\left( {1 + {\alpha ^2}} \right)}^{n + 1}}}}} \ ,$$
where  ${\alpha ^2}$
 is the mean photon number of Alice's mode. We note that $\alpha^2={\rm sinh}^2r$, where $r$ is the squeezing parameter of the two-mode squeezing operator $$S\left( \xi  \right) = \exp \left( {\xi {\hat a} {\hat b}-\xi{\hat a}^\dag{\hat b}^\dag } \right), \ \xi  = r{e^{i\theta} } \ ,$$ where $\theta$ represents the orientation of the squeezing, and where ${\hat a}$ and ${\hat a^\dagger }$  represent the  annihilation and creation operators, respectively, of mode $A$. Here, we assume $\theta=0$.

 \hfil

\noindent\textbf{Result 1:} The quantum state after the channel can be written as
\[\begin{array}{*{20}{l}}
{{\left| \psi  \right\rangle }_{TPS}}
=&- \frac{1}{{\sqrt {{P_1}} }}\sum\limits_{n = 1}^\infty  {\sum\limits_{k = 0}^{n - 1} {\sum\limits_{m = 0}^\infty  {\sum\limits_{l = 0}^m {s_{n,k,m,l}} } } } \\
&{\times {{\left| {n,n - 1 - k + l,k + m - l,m} \right\rangle }_{A{B_2}EF}},}\\
\end{array}\]
 where
 ${s_{n,k,m,l}} = {\alpha _n}{\beta _m}{( - 1)^k}r_{n,1}^{T_S}r_{n - 1,k}^{{T_E}}r_{m,l}^{{T_E}}{z_{n - 1,k,m,l}}$, and the other variables introduced above are defined in the following proof.

\noindent\textbf{Proof:} Initially we have the following description of the combined $AB_0C_0B_1C$ mode
\[\begin{array}{*{20}{l}}
 {\left| \psi  \right\rangle _{A{B_0}{C_0}{B_1}C}} &= \sum\limits_{n = 0}^\infty  {{\alpha _n}} {\left| {n,n} \right\rangle _{A{B_0}}}{\left| {0,0,0} \right\rangle _{{C_0}{B_1}C}} \\
  &= \sum\limits_{n = 0}^\infty  {{\alpha _n}} \frac{{{{\left( {\hat b_0^\dag } \right)}^n}}}{{\sqrt {n!} }}{\left| {n,0} \right\rangle _{A{B_0}}}{\left| {0,0,0} \right\rangle _{{C_0}{B_1}C}} \ .
  \end{array}\]
  The\ presence\ of\ the\ beam-splitter\ at\ the\ PS stage\ alters\ this\ combined\ mode\ to\ the\ form
  \[\begin{array}{*{20}{l}}
  &\sum\limits_{n = 0}^\infty  {{\alpha _n}} \frac{{{{(\sqrt {{T_S}} \hat b_1^\dag  - \sqrt {1 - {T_S}} {{\hat c}^\dag })}^n}}}{{\sqrt {n!} }}{\left| {n,0} \right\rangle _{A{B_0}}}{\left| {0,0,0} \right\rangle _{{C_0}{B_1}C}} \\
  = &\sum\limits_{n = 0}^\infty  {{\alpha _n}} \sum\limits_{k = 0}^n {{{( - 1)}^k}r_{n,k}^{{T_S}}} {\left| {n,0} \right\rangle _{A{B_0}}}{\left| {0,n - k,k} \right\rangle _{{C_0}{B_1}C}} \ ,
 \end{array}\]
 where
 $r_{n,k}^T = \sqrt {\left( {\begin{array}{*{20}{c}}
   n  \\
   k  \\
\end{array}} \right)} {(\sqrt T) ^{n - k}}{\sqrt {1 - T} ^k}$.
We assume that the subtraction is for the single photon case (i.e. $k = 1$
  and $C = \left| 1 \right\rangle $). Tracing out mode ${B_0}$,
$C$,   and ${C_0}$
  we have,
  \[{\left| \psi  \right\rangle _{A{B_1}}} =  - \frac{1}{{\sqrt {{P_1}} }}\sum\limits_{n = 1}^\infty  {{\alpha _n}r_{n,1}^{T_S}} {\left| {n,n - 1} \right\rangle _{A{B_1}}},\]
where
$$P_1=\sum\limits_{n = 1}^\infty \left( {{\alpha_n}r_{n,1}^{T_S}}\right )^2$$
 is the probability of subtracting one photon.
Similar to Alice, Eve's initial TMSV state is,
\[{\left| \psi  \right\rangle _{{E_0}F}} = \sum\limits_{m = 0}^\infty  {{\beta _m}{{\left| {m,m} \right\rangle }_{{E_0}F}}} \]
with
$${\beta _m} = \sqrt {\frac{{{\beta ^{2m}}}}{{{{\left( {1 + {\beta ^2}} \right)}^{m + 1}}}}} \ ,$$
where ${\beta ^2}$
  is the mean photon number of  Eve's mode - a parameter used to simulate the channel noise.

As it passes the channel, mode  ${B_1}$
 evolves to mode  ${B_2}$. Prior to Eve acting on the incoming states we have the following description of the combined $AB_1E_0EFB_2$ mode
\[\begin{array}{*{20}{l}}
 {\left| \psi  \right\rangle _{A{B_1}{E_0}EF{B_2}}}
  =&  - \frac{1}{{\sqrt {{P_1}} }}\sum\limits_{n = 1}^\infty  {{\alpha _n}r_{n,1}^{T_S}} {\left| {n,n - 1} \right\rangle _{A{B_1}}} \\
  &\otimes \sum\limits_{m = 0}^\infty  {{\beta _m}{{\left| {m,m} \right\rangle }_{{E_0}F}}} {\left| {0,0} \right\rangle _{{B_2}E}} \ .
\end{array}\]

The presence of the beam-splitter at Eve  alters this combined mode to the form
  \[\begin{array}{c}
   - \frac{1}{{\sqrt {{P_1}} }}\sum\limits_{n = 1}^\infty  {{\alpha _n}r_{n,1}^{T_S}} \frac{{{{(\sqrt {{T_E}} \hat b_2^\dag  - \sqrt {1 - {T_E}} {{\hat e}^\dag })}^{n - 1}}}}{{\sqrt {(n - 1)!} }}{\left| {n,0} \right\rangle _{A{B_1}}} \\
  \otimes \sum\limits_{m = 0}^\infty  {{\beta _m}\frac{{{{(\sqrt {{T_E}} \hat e^\dag  + \sqrt {1 - {T_E}} {{\hat b_2}^\dag })}^m}}}{{\sqrt {m!} }}{{\left| {0,m} \right\rangle }_{{E_0}F}}} {\left| {0,0} \right\rangle _{{B_2}E}} \\
  =  - \frac{1}{{\sqrt {{P_1}} }}\sum\limits_{n = 1}^\infty  {{\alpha _n}r_{n,1}^{T_S}} \sum\limits_{k = 0}^{n - 1} {{{( - 1)}^k}r_{n - 1,k}^{{T_E}}}  \\
  \times \sum\limits_{m = 0}^\infty  {{\beta _m}} \sum\limits_{l = 0}^m {r_{m,l}^{{T_E}}{z_{n - 1,k,m,l}}  }  \\
  \times {\left| {n,n - 1 - k + l,k + m - l,m,0,0} \right\rangle _{A{B_2}EF{B_1}{E_0}}} \ , \\
 \end{array}\]
 where
 $${z_{n,k,m,l}} = \sqrt {\left( {\begin{array}{*{20}{c}}
   {n - k + l}  \\
   l  \\
\end{array}} \right)} \sqrt {\left( {\begin{array}{*{20}{c}}
   {k + m - l}  \\
   k  \\
\end{array}} \right)} \ . $$
Rearranging the summation and tracing out  ${B_1}$
 and ${E_0}$
  we arrive at the Result~1.

\section {Photon subtraction at receiver side}
\begin{figure}[h]
	\includegraphics[width=0.5\textwidth]{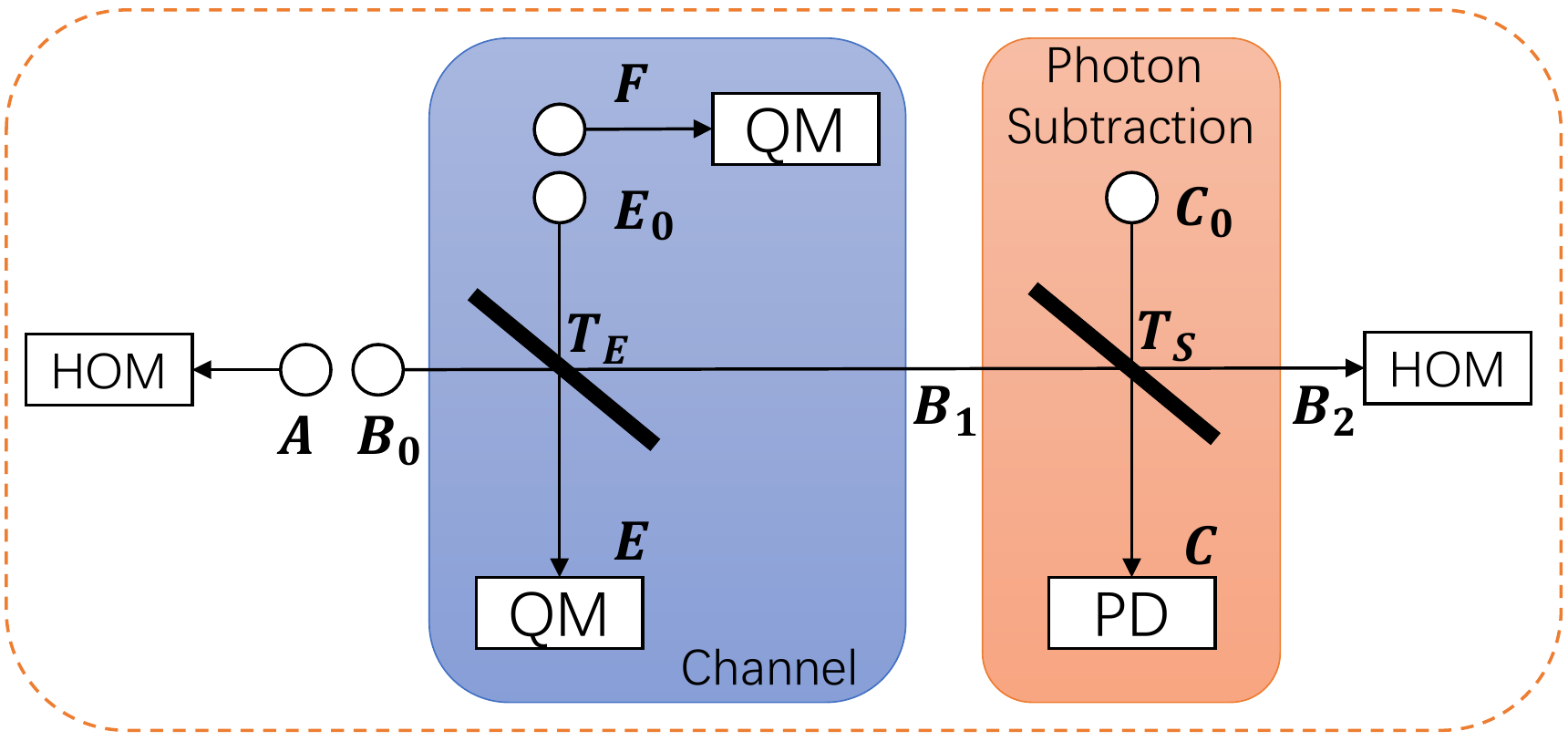}
	\caption{Photon subtraction at receiver side (R-PS).
 Here Alice (ground station) prepares a TMSV ($A_0-B_0$),
 sending $B_0$ through a channel controlled by Eve using a  beam-splitter with transmissivity $T_E$.
  The exiting mode $B_1$ is sent by Eve to Bob (the satellite) who undertakes a PS process on $B_1$ using a beam-splitter with transmissivity $T_S$, leading to $B_2$.\label{RPS}}
\end{figure}

If the photon subtraction occurs at the receiver side instead of the  transmitter side (Fig.~\ref{RPS}), a  different outcome is achieved for the final state -  a result previously  derived in \cite{5n}. We simply provide that result here (the proof follows a similar path to that given for PS at the transmitter). However, we note the work of \cite{5n} considers the fixed-attenuation channel only, and therefore the results of that work cannot be directly utilized for the Earth-satellite channels we are concerned with here.

Prior to the PS at the receiver the quantum state is given by

\[\begin{array}{*{20}{l}}
 {\left| \psi  \right\rangle _{A{B_1}EF}} =& \sum\limits_{n = 0}^\infty  {\sum\limits_{k = 0}^n {\sum\limits_{m = 0}^\infty  {\sum\limits_{l = 0}^m {{\alpha _n}{\beta _m}{{( - 1)}^k}r_{n,k}^{{T_E}}}r_{m,l}^{{T_E}}{z_{n,k,m,l}} } } }  \\
  &\times {\left| {n,n - k + l,k + m - l,m} \right\rangle _{A{B_1}EF}} \ . \\
 \end{array}\]
After the channel, Bob performs PS on $B_1$, leading to the $B_2$ mode. This latter mode is subsequently used in Bob's
   homodyne detection.\hfil

\noindent{\textbf{Result 2}:}   The photon subtracted quantum state at the receiver can be written
\[\begin{array}{*{20}{l}}
{{\left| \psi  \right\rangle }_{RPS}}
=&- \frac{1}{{\sqrt {{P'_1}} }}\sum\limits_{n = 0}^\infty  {\sum\limits_{k = 0}^{n } {\sum\limits_{m = 0}^\infty  {\sum\limits_{l = 0}^m {s'_{n,k,m,l}} } } } \\
&{\times {{\left| {n,n - 1 - k + l,k + m - l,m} \right\rangle }_{A{B_2}EF}},}\\
\end{array}\]
where ${s'_{n,k,m,l}} = {\alpha _n}{\beta _m}{( - 1)^k}r_{n - k + l,1}^{{T_S}}r_{n,k}^{{T_E}}r_{m,l}^{{T_E}}{z_{n,k,m,l}}$
and $P'_1$ is a new normalization constant (\emph{cf}. Eq. (19) of \cite{5n}).

\section {Performance analysis}

\subsection {Covariance Matrix}

Before moving into our investigation of the secret key rate we note that the covariance matrix of a given state $\left| \psi  \right\rangle_{AB}$
  with two modes $A$  and mode $B$, can be written as
$${{\bf{M}}_{{\bf{AB}}}} = \left[ {\begin{array}{*{20}{c}}
   {{V_A}{\bf{I}}} & {{C_{AB}}{\bf{\sigma }}}  \\
   {{C_{AB}}{\bf{\sigma }}} & {{V_B}{\bf{I}}}  \\
\end{array}} \right]\ , $$
where ${\bf{I}} = diag(1,1)$, ${\bf{\sigma }} = diag(1, - 1)$. Here, \[{V_A} = \left\langle \psi  \right|1 + 2{\hat a^\dag }\hat a\left| \psi  \right\rangle_{AB} \]
is the variance of mode  $A$ (likewise $V_B$), and \[{C_{AB}} = \left\langle \psi  \right|\hat a\hat b + {\hat a^\dag }{\hat b^\dag }\left| \psi  \right\rangle_{AB} \]
is the covariance between mode $A$  and mode  $B$.

Consider next the  variances of mode $A$ and mode $F$ following PS at the transmitter. Using the above, we can see that the variances of mode $A$ and  $F$ can be given as,
\[\begin{array}{*{5}{l}}
 {V_A}
 &=  \left\langle \psi  \right|1 + 2{{\hat a}^\dag }\hat a{\left| \psi  \right\rangle _{TPS}} \\
 &=   1 - \frac{2}{{\sqrt {{P_1}} }}\left\langle \psi  \right|\sum\limits_{n = 1}^\infty  {\sum\limits_{k = 0}^{n - 1} {\sum\limits_{m = 0}^\infty  {\sum\limits_{l = 0}^m {n  {s_{n,k,m,l}}} } } } \\
 &\ \ \ \times {\left| {n,n - 1 - k + l,k + m - l,m} \right\rangle _{A{B_2}EF}} \ ,
 \end{array}\]
\[\begin{array}{*{5}{l}}
 {V_F}
 &= \left\langle \psi  \right|1 + 2{{\hat f}^\dag }\hat f{\left| \psi  \right\rangle _{TPS}} \\
 &= 1 - \frac{2}{{\sqrt {{P_1}} }}\left\langle \psi  \right|\sum\limits_{n = 1}^\infty  {\sum\limits_{k = 0}^{n - 1} {\sum\limits_{m = 0}^\infty  {\sum\limits_{l = 0}^m {m  {s_{n,k,m,l}}} } } }  \\
 &\ \ \  \times {\left| {n,n - 1 - k + l,k + m - l,m} \right\rangle _{A{B_2}EF}} \ , \\
 \end{array}\ \]
respectively.
Likewise, the  covariance between two different modes, say $E$ and $F$, can be given by
\[\begin{array}{*{20}{l}}
 {C_{EF}} &= \left\langle \psi  \right|\hat e\hat f + {{\hat e}^\dag }{{\hat f}^\dag }{\left| \psi  \right\rangle _{TPS}} \\
  &=  - \frac{1}{{\sqrt {{P_1}} }}\left\langle \psi  \right|\left. \varphi  \right\rangle\ \ ,
   \end{array}\]
 where
  \[\begin{array}{c}
 \left| \varphi  \right\rangle  = \sum\limits_{n = 1}^\infty  {\sum\limits_{k = 0}^{n - 1} {\sum\limits_{m = 0}^\infty  {\sum\limits_{l = 0}^m {{s_{n,k,m,l}}\sqrt {m + 1} \sqrt {k + m - l + 1} } } } }  \\
  \times {\left| {n,n - 1 - k + l,k + m - l + 1,m + 1} \right\rangle _{A{B_2}EF}} \ +  \\
 \sum\limits_{n' = 1}^\infty  {\sum\limits_{k' = 0}^{n' - 1} {\sum\limits_{m' = 1}^\infty  {\sum\limits_{l' = 0}^{m'-1} {{s_{n',k',m',l'}}\sqrt {m'} \sqrt {k' + m' - l'} } } } }  \\
  \times {\left| {n',n' - 1 - k' + l',k' + m' - l' - 1,m' - 1} \right\rangle _{A{B_2}EF}} \ . \\
 \end{array}\]

 Similar variance and covariance terms can be derived for PS at the receiver. These terms can be calculated numerically simply by using the fact that $\left\langle{n,k,m,l}|{n',k',m',l'}\right\rangle = \delta_{nkml,n'k'm'l'}$. The usefulness of such terms will become evident when we calculate the keys rates, an issue we turn to next.

\subsection{The Secret Key Rate}

Under a collective attack, the key rate is related to the difference of $I(A:{B_2})$ - the mutual information between mode $A$  and mode $B_2$; and $\chi({B_2}:EF)$ - the Holevo information that Eve can extract from her measurement \cite{Weedbrook2012}. More specifically, we can say, the key rate (per pulse generated by the source laser) is,
$$K({T_E}) = {P}\left[ {fI(A:{B_2}) - \chi ({B_2}:EF)} \right] \ ,$$
where $f$  is the decoding reconciliation efficiency, and  $P$ is the probability of subtracting one photon in the PS.
However,  calculation of the key rate for a non-Gaussian state  is analytically not tractable since the non-Gaussian state has more than two non-zero moments. To make progress, we utilize the Gaussian state (metrics of which will be indicated by the subscript $G$) that produces the same covariance matrix ${\bf{{ M}}}$
  as the non-Gaussian state ${\left| \psi  \right\rangle _{A{B_2}EF}}$. This provides a lower bound for the key rate by the theorem of Gaussian optimality \cite{2n}. Emphasizing that all key rates  discussed from this point on are bounds, we have\footnote{Note, the beam-splitter attack we use is the most pragmatic, but it is slightly sub-optimal. Under an optimal attack (purification), the key rate will be approximately 1.1dB lower for all our schemes.}
\[K({T_E}) \ge {P}\left[ {f{I_G}(A:{B_2}) - {\chi _G}({B_2}:EF)} \right]\ , \]
where \cite{Weedbrook2012}
\[{I_G}(A:{B_2}) = \frac{1}{2}{\log _2}\frac{{{V_{{B_2}}}}}{{{V_{{B_2}|A}}}} \ , \]
and the conditional variance ${V_{{B_2}|A}}$ is
\[{V_{{B_2}|A}} = {V_{{B_2}}} - \frac{{{C_{A{B_2}}}^2}}{{{V_A}}} \ .\]
For Eve's stolen information, we can write
\[{\chi _G}({B_2}:EF) = \sum\limits_i {g(v_i^{EF})}  - \sum\limits_j {g(v_j^{EF|{B_2}})} \ , \]
where
$$g(v) = \frac{{v + 1}}{2}{\log _2}\frac{{v + 1}}{2} - \frac{{v - 1}}{2}{\log _2}\frac{{v - 1}}{2} \ .$$ In the above,
 ${v^{EF}}$
 and  ${v^{EF|{B_2}}}$
 are the symplectic eigenvalues of the covariance matrices ${{\bf{M}}_{{\bf{EF}}}}$
  and ${{\bf{M}}_{{\bf{EF|}}{{\bf{B}}_{\bf{2}}}}}$, respectively,
  where \cite{Weedbrook2012}
\[{{\bf{M}}_{{\bf{EF|}}{{\bf{B}}_{\bf{2}}}}} = {{\bf{M}}_{{\bf{EF}}}} - \left[ {\begin{array}{*{20}{c}}
   {{C_{E{B_2}}}{\bf{I}}}  \\
   {{C_{F{B_2}}}{\bf{\sigma }}}  \\
\end{array}} \right]\left[ {\begin{array}{*{20}{c}}
   {{V_{{B_2}}}^{ - 1}} & 0  \\
   0 & 0  \\
\end{array}} \right]{\left[ {\begin{array}{*{20}{c}}
   {{C_{E{B_2}}}{\bf{I}}}  \\
   {{C_{F{B_2}}}{\bf{\sigma }}}  \\
\end{array}} \right]^T} \ . \]
Finally, we can now determine the bound on the key rate achieved in the satellite lossy channel by taking the average over all possible transmission coefficient values, namely,
${K_{avg}} = \int {p({\eta})K({\eta^2})d{\eta}}$. Allowing the initial squeezing to be dependent on $\eta$ allows for further optimization of the key rate - an issue we ignore for simplicity.
\section{Simulation results}
\begin{figure}[h]
	\includegraphics[width=0.38\textwidth]{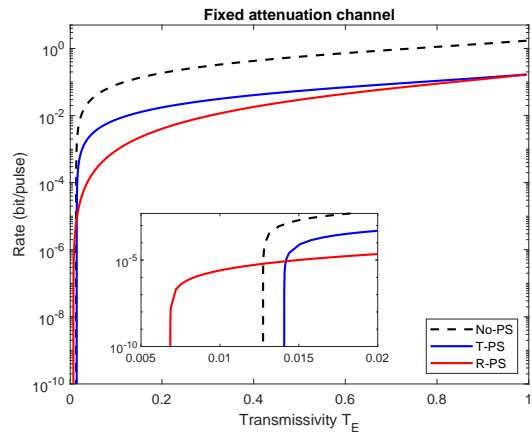}
	\centering
	\caption{The key  rate vs. transmissivity. \label{fixeda}}
\end{figure}
\begin{figure}[h]
	\includegraphics[width=0.38\textwidth]{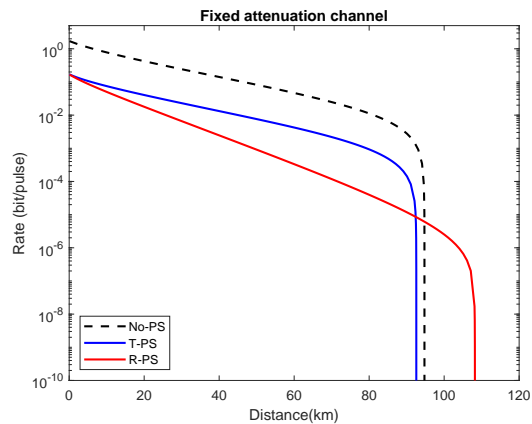}
	\centering
	\caption{The key  rate vs. distance. \label{fixedb}}
\end{figure}
 \begin{figure}[h]
	\includegraphics[width=0.4\textwidth]{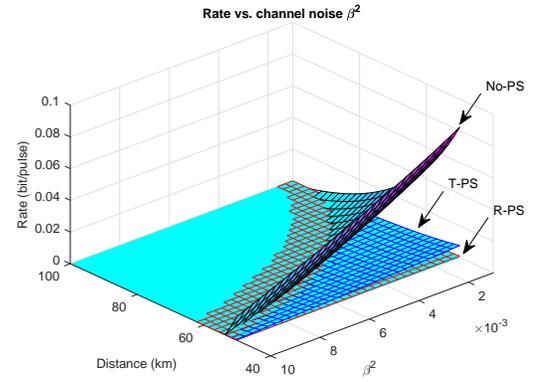}
	\centering
	\caption{The  key rate  over the fixed channel for different noise conditions. The top, middle, and bottom layers are No PS, T-PS and R-PS, respectively. \label{sat3}}
\end{figure}
\begin{figure}[h]
	\includegraphics[width=0.4\textwidth]{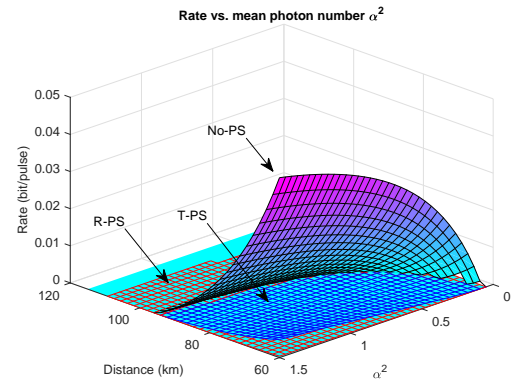}
	\centering
	\caption{The  key rate  over the fixed channel for different mean photon number. The top, middle, and bottom layers are No PS, T-PS and R-PS, respectively. \label{sat4}}
\end{figure}
 \begin{figure}[h]
	\includegraphics[width=0.38\textwidth]{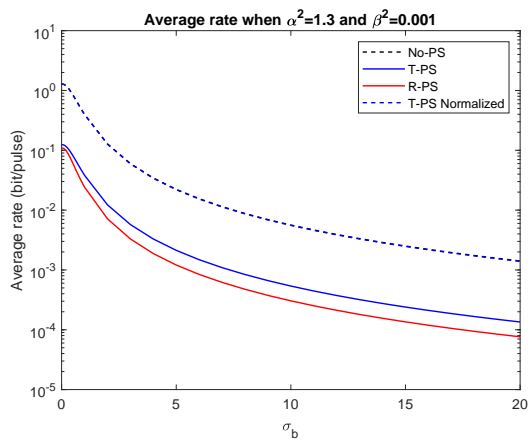}
	\centering
	\caption{The  key rate averaged over the satellite channel as a function of the standard deviation of the beam wandering for range 0-20. \label{sat}}
\end{figure}
\begin{figure}[h]
	\includegraphics[width=0.45\textwidth]{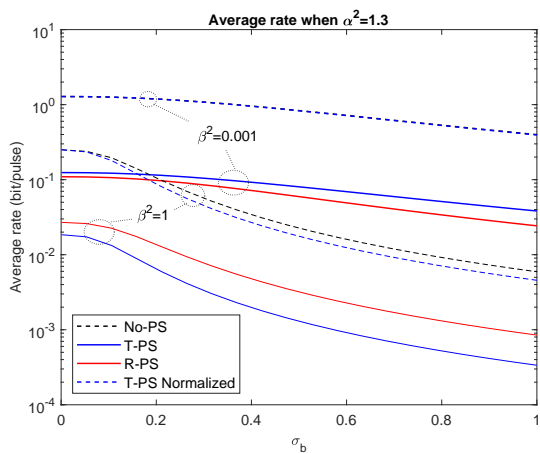}
	\centering
	\caption{A close up of the  key rate averaged over the satellite channel as a function of the standard deviation of the beam wandering for the range 0-1. \label{sat2}}
\end{figure}
%
For comparison purposes we first consider a non-variable attenuation channel, before comparing the performance of our three schemes for the satellite channel we have discussed earlier in the paper.
Unless otherwise stated, the parameters utilized  in the calculations shown are ${\alpha ^2} = 1.3$, ${\beta ^2} = 0.001$, $f = 0.95$, and $T_S = 0.9$ (for simplicity a detector efficiency of 1 is assumed). The infinite summation limits are constrained to 20 for $n$ and $m$ \cite{errorv1}.

%

 As stated, we first consider a fixed attenuation channel. Here we fix the value of $\alpha^2$ for all attenuation conditions. We plot the key rate against transmissivity in Fig.~(\ref{fixeda}), and against distance in Fig.~(\ref{fixedb}).
In Fig.~(\ref{fixedb}) we assume that the channel has a fixed attenuation of $0.2$dB/km. The results of Figs.~(\ref{fixeda})-(\ref{fixedb}) show that the R-PS scheme has the longest key distribution range at a cost of a reduced key rate. That is, the R-PS scheme is in some sense the most robust against  channel attenuation (provides a non-zero key rate at the largest distance).
 We further compare the performance of the three schemes as a function of the noise ${\beta ^2}$
  and the mean photon number ${\alpha ^2}$ (i.e. sinh$^2r$, $r$ being the squeezing parameter) - the results of which are shown in Figs.~(\ref{sat3}) and (\ref{sat4}), respectively. Note, that in these figures the rates are not plotted in the logarithmic domain so the comparison in the small rate region is not as apparent. As can be seen, for some parameter space we  find distances where the T-PS scheme shows better key rate performance than the other schemes. We also find the T-PS and R-PS schemes can outperform the No-PS scheme in some parameter space (again we caution that optimisation of the initial squeezing can alter these conclusions).

We next investigate the key rates of the three schemes in the variable Earth-satellite channel, calculating their average key rates under different average channel fluctuations, quantified using   ${\sigma _b}$ within equations~(\ref{f1})-(\ref{f2}). These results are shown in Fig.~(\ref{sat}).
 The No-PS case shows  better performance in terms of key rate for the entire range of channel conditions - a result not found for the fixed attenuation case. The PS cases (T-PS and R-PS) are impacted by the low probability of obtaining a subtracted photon in any given pulse, and this effect dominates when channel averaging over the fading channel is accounted for. The blue dashed curve (marked normalized) in Fig.~(\ref{sat}) show the impact of a quantum memory in place such that the low probability for PS can be negated. Here the schemes are assumed to be \emph{a priori} storing the required states in memory, then sending the same rate of quantum states into the satellite channel on-demand.
 A close up at low $\sigma_b$ is shown in  Fig.~(\ref{sat2}) for different noise conditions. These latter results show the rates possible in very-high quality downlinks from the satellite-to-Earth.\footnote{ Note that $\sigma_{b}=1$ corresponds to approximately 5dB of loss. Such low loss rates are possible for well-engineered systems in which diffraction of the beam is the major factor contributing to photon loss.}

 A main aim of our study was to determine whether PS at the transmitter-side outperforms PS at the receiver-side for a range of Earth-Satellite channels (where no instantaneous channel-dependent optimisation of squeezing occurs at the transmitter). Figs.~(\ref{sat})-(\ref{sat2}) provide an answer to this question -  yes. This result holds for all anticipated channel conditions (only at unrealistic noise levels is the opposite found).

\section{Conclusions}
We have studied the use of non-Gaussian CV quantum states - created via photon subtraction - in the context of a  straightforward QKD protocol.
More specifically, we have studied the lower-bounds on secret key rates delivered by such states.
Contrary to what is found in fixed attenuation channels (such as optical fiber), we find that for the variable-channels anticipated for Earth-satellite communications, photon subtraction at the transmitter, for an initially fixed squeezing,  outperforms photon subtraction at the receiver for all realistic conditions.
The authors acknowledge  support  from the UNSW,  the CSC, and Northrop Grumman.


\end{document}